\documentstyle[12pt,sprocl]{article}
\linethickness{0.2mm}
\begin{document}
\begin{center}
{\Large\bf The Role of Renormalization Group }
 \\  {\Large\bf in Fundamental Theoretical Physics}\footnote{
The text of the talk presented at the Conference "RG-96" (Dubna, Aug 96).
To appear at the proceedings.}   \\
\medskip

{\large\sf Dmitri V. SHIRKOV} \\                       
\medskip
{\small\it Bogoliubov Theor. Lab. JINR, Dubna, RUSSIA \\
 e-mail: shirkovd@thsun1.jinr.dubna.su }
 \end{center}
\bigskip

\vspace{5mm}
\noindent{\large\bf 1. Introduction (Logic of science)} \\ 

Here, I would like to discuss some general aspects of the logical
structure of modern fundamental science and, in particular, the  place
and role of \underline{r}enormalization \underline{g}roup (RG) in it.
Here, by RG we mean the Stueckelberg--Bogoliubov formulation of the
Renormalization Group, that is a one-parameter continuous group in a
usual mathematical sense. \par
    The importance of symmetries and groups in
fundamental theoretical physics have been realized by some of the leading
theorists more than half a century ago. One of its most prominent
advocates, Eugene Wigner, proposed a hierarchical scheme establishing
a relation between three categories: "symmetry or invariance principles",
"laws of nature" and "events".    \par
As he wrote in 1964 (see pp. 38 and 30 in Ref. \cite{wig}) :
\begin{quote}
 "What I would like to discuss instead is the general role of symmetry
and invariance principles in physics, both modern and classical. More
precisely, I would like to discuss the relation between three categories
which play a fundamental role in all natural sciences: events, which
are the raw material for the second category, the laws of nature, and
symmetry principles for which I would like to support the thesis that
the laws of nature form the raw material." 

"... the progression from events to laws of nature, and from laws of
nature to symmetry or invariance principles, is what I meant by the
heirarchy of our knowledge of the world around us." 
\end{quote}

\vspace{1mm}
	\begin{figure}[hpt]
        \begin{center}
\fbox{
        \vbox to 9mm {

\vspace{2.3mm}
\centerline{Physical ~$ \fbox{\sf EVENTS}~ \Rightarrow ~\framebox{\sf LAWS}~
\mbox{of Nature} ~\Rightarrow ~$ \fbox{\sf PRINCIPLES}~ of Symmetry}
}                      }
          \end{center}   
	  \caption{
{\bf  WIGNER Hierarchy}: {\sl Events form the basis for laws.
Laws provide the raw material for principles.}
			        }
	  \label{fig1}
          \end{figure}   

This hierarchy follows the line of "science construction", of extracting
regularities from observation, regularities (laws and principles) that
form the skeleton of physical science. \par
 However, principles and laws obey predictive ability. To follow the
inner logic of science one should proceed in the opposite direction.
Again, according to Wigner (p. 17 in Ref.~\cite{wig}):
\begin{quote}
"... the function of the invariance principles to provide a structure
or coherence to the laws of nature just as the laws of nature provide
a structure and coherence to the set of events." 
\end{quote}
This quotation with some details added can be visualized in the
form of a scheme (see Ref. \cite{gross}):

	\begin{figure}[hptb]
        \begin{center}
\begin{picture}(140,80)
\put(0,0){\line(1,0){140}}
\put(0,80){\line(1,0){140}}
\put(0,0){\line(0,1){80}}
\put(140,0){\line(0,1){80}}
\put(50,72){\fbox{\sf PRINCIPLES} of Symmetry}
\put(5,64){\underline{Couplig constants}}
\put(88,64){\underline{Representations, ...}}
\put(57,42){\fbox{\sf LAWS} of Nature}
\put(88,34){\underline{Initial Conditions}}
\put(39,4){Physical \fbox{\sf EVENTS} }
\put(63,70){\vector(0,-1){23}}
\put(63,40){\vector(0,-1){31}}
\put(20,60){\vector(3,-1){37}}
\put(104,60){\vector(-3,-1){36}}
\put(104,31){\vector(-3,-2){31}}
\end{picture}
          \end{center}
	 \caption{
	 {\bf Logic of science by E. Wigner:}
	 {\sl Principles provide a structure to the laws of nature. 
	 Laws define the Events pursuit.
			        }     }
	  \label{fig2}
          \end{figure}
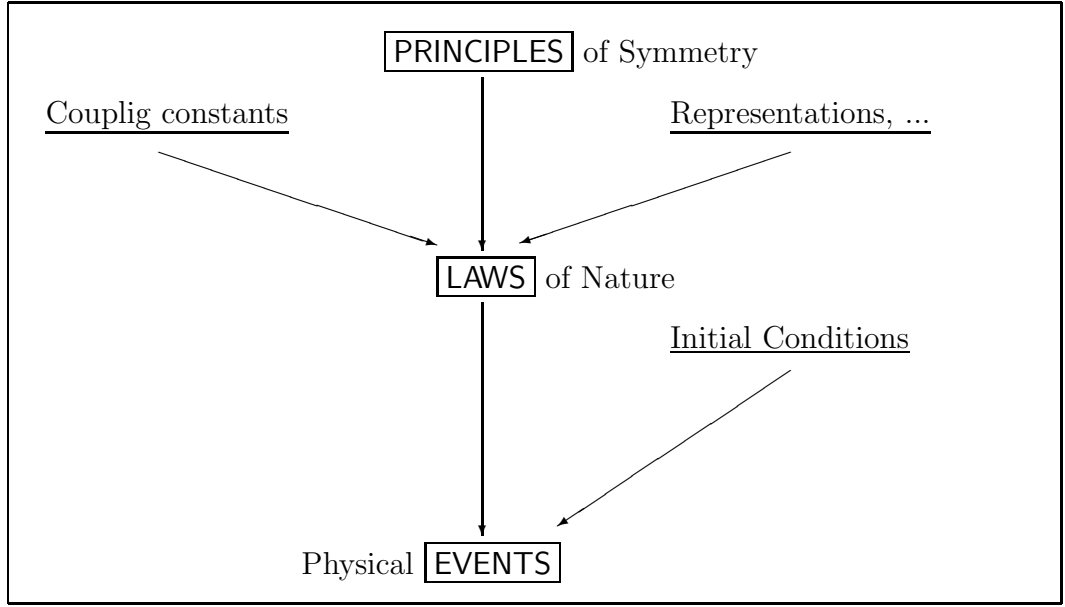

 In what follows, we would like to discuss the validity of the Wigner
 scheme in modern physical science and indicate the place of RG in it.
 However, for this purpose, the scheme, Fig.2, is a bit sketchy. We have
to modify it. Our first comment relates to the category of 'principles'. \\
\medskip

\noindent{\large\bf 2. Comment on "Principles"} \\

Wigner paid attention mainly to principles of symmetry, like
space-time (Poincar\'e Invariance, P, T) and internal symmetries
 (Isospin $\to$ Flavours, Colour). Meanwhile, in fundamental physics
we deal also with some other principles, {\sf General Principles},
like:
\medskip \par
-- {\sf Principle of QUANTUM PRIORITY} which states that
"quantum level of nature is the basic one and classical physics is
 secondary, being the limiting case of a quantum picture";
\smallskip \par
-- {\sf Principle of UNITARITY} that reflects the
"conservation of probability" ;
\medskip \par
-- {\sf Principle of CAUSALITY}: "Future cannot influence the past"
(related to the mistery of a 'Time arrow');
\medskip \par
-- {\sf Principle of RENORMALIZABILITY} \footnote{This needs quantum
notions to be formulated in detail.} that acts as a selection rule
for QFT models and can be formulated~\cite{prais,heis} as follows :
"The given model of field interaction should be realizable on the
quantum level".
{\small
 In combination with the principle of quantum priority this means
that the renormalizability property should be considered as a
nessesary condition for a given QFT model to have a chance to
describe the Nature: i.e., {\sf RENORMALIZABILITY = RELIABILITY.}}
\medskip \par
-- The {\sf GAUGE DYNAMICS Principle} \footnote{This needs
quantum notions to be motivated.} stating that the form of a dynamics,
of a field interaction, should be deduced from a symmetry (by its
"localizing").        

\vspace{1cm}
\noindent{\large\bf 3. Are "Equations" equivalent to "Laws"?}  \\

 The second comment relates to the Wigner's category "laws of nature".
 In our opinion they, generally, should not be identified with  {\it
Dynamical Equations} deductible from some basic Principles. Rather,
these "Laws" are to be related to {\it Solutions} of Dynamical
Eqs. To illustrate, consider the case of the Standard Model (SM) in QFT.
\par
    The most fashionable current topics in SM (Grand Unification,
SuSy generalization,
quantum gravity,...) are related to 'extremely high energy region'.
However, there are two issues lying in the experimentally studied
domain. These are: "Confinement in QCD" and "Vector boson masses
in the ElectroWeak Theory": \par
-- All experts agree that we have correct QCD equations responsible
for strong interaction. However, the confinement phenomenon, being
an essentially nonlinear quantum effect, still is not understood. \par
-- The origin of gauge $W^{\pm}$ and $Z^0$ boson masses is "explained"
by the so-called 'Higgs mechanism'. It is highly artificial and,
technically, is based upon a very specific scalar field with
imaginary mass and quartic self-interaction. ({\small This mechanism
also predicts particles which have not been observed yet.}) The scalar
field introduction destroys the whole beauty of the Gauge Dynamics
principle. Meanwhile, there are serious reasons~\cite{ew-col} to
believe that spontaneos symmetry breaking of gauge symmetry can be
treated as an intrinsic feature of non-Abilean quantum
gluon field, as a nonlinear quantum phenomenon. \par
  Both issues are related to common non-linear quantum topic --
\medskip

\centerline{\sf Structure of the ground state of non-Abelian~Quantum Field.}
\smallskip
Here, we have beautiful equations, like the QCD ones,
whose structure is determined by principles. However, we are unable
to extract from them the very basic feature of strong interaction
(confinement of coloured objects) and some other important information
related to experiment. Instead, for the latter purpose we have to be
satisfied with effective semi-phenomenological model constructions,
like "MIT bag","Dubna bag", "low-energy chiral model" which are not
directly related to general and symmetry principles. This means that
"Equations", in practice, do not define the essential features of
a system. It is improper to treat them as "Laws".
\smallskip \par
Another illustration is provided by the history of superconductivity.
This phenomenon stayed unsolved by theorists for about 45 years. In
the course of the first 15 years there was no adequate theoretical
basement (of quantum mechanics = QM). However, during subsequent
three decades we had general belief that superconductivity had to be
understood as a macroscopical QM effect, but not understanding of this
phenomenon on the basis of QM description of electron gas in metal.
Instead, we used to content ourselves with semi-phenomelogical
constructions like that of Londons and Ginzburg--Landau. Just they
appeared as {\it "laws ... providing a structure to the set of
events".}\cite{bcs}  The situation is pictured on Fig.3.

	\begin{figure}[h]
        \begin{center}
\fbox{
        \vbox to 17mm {
\par
$ 1911 ~~ \to~~ 15~years ~~ \to~~ 1926/7 \hspace{3mm}~\to \to
~~{\bf 30~years !}~~~ \to \to \hspace{4mm} 1956/7 $

\vspace{2mm}
\par    Experiment \hspace{22mm} "Theory" \hspace{46mm} "Laws"=
\par    Discovery \hspace{27mm} QM Eqs.  \hspace{46mm} Solution
}                  }
          \end{center}
	  \caption{ \sl
History of superconductivity (Difference between "Equations" and "Laws").
			        }
	  \label{fig3}
          \end{figure}

\vspace{1mm}
 Thus, we arrive at the conclusion that "Laws" should be
substituted by two notions : \par
-- {\sf Equations} that can be deduced from Principles; \par
-- {\sf Solutions} of equations that are equivalent to "Laws" in a
sense that they determine the physical system behaviour. In short --

\centerline{\it Between "Principles" and "Laws" there should stand
"Equations".}                                            \par
 This means that principles provide the structure and coherence just
to dynamical equations which, in a sense, could be treated as laws of
nature formulated in a general form. However, as a rule, in modern
science they have no close relation to events. These are rather
solutions to equations which provide structure and coherence to sets
of physical events. Instead of the Wigner-like scheme, Fig.2, we get: \\

	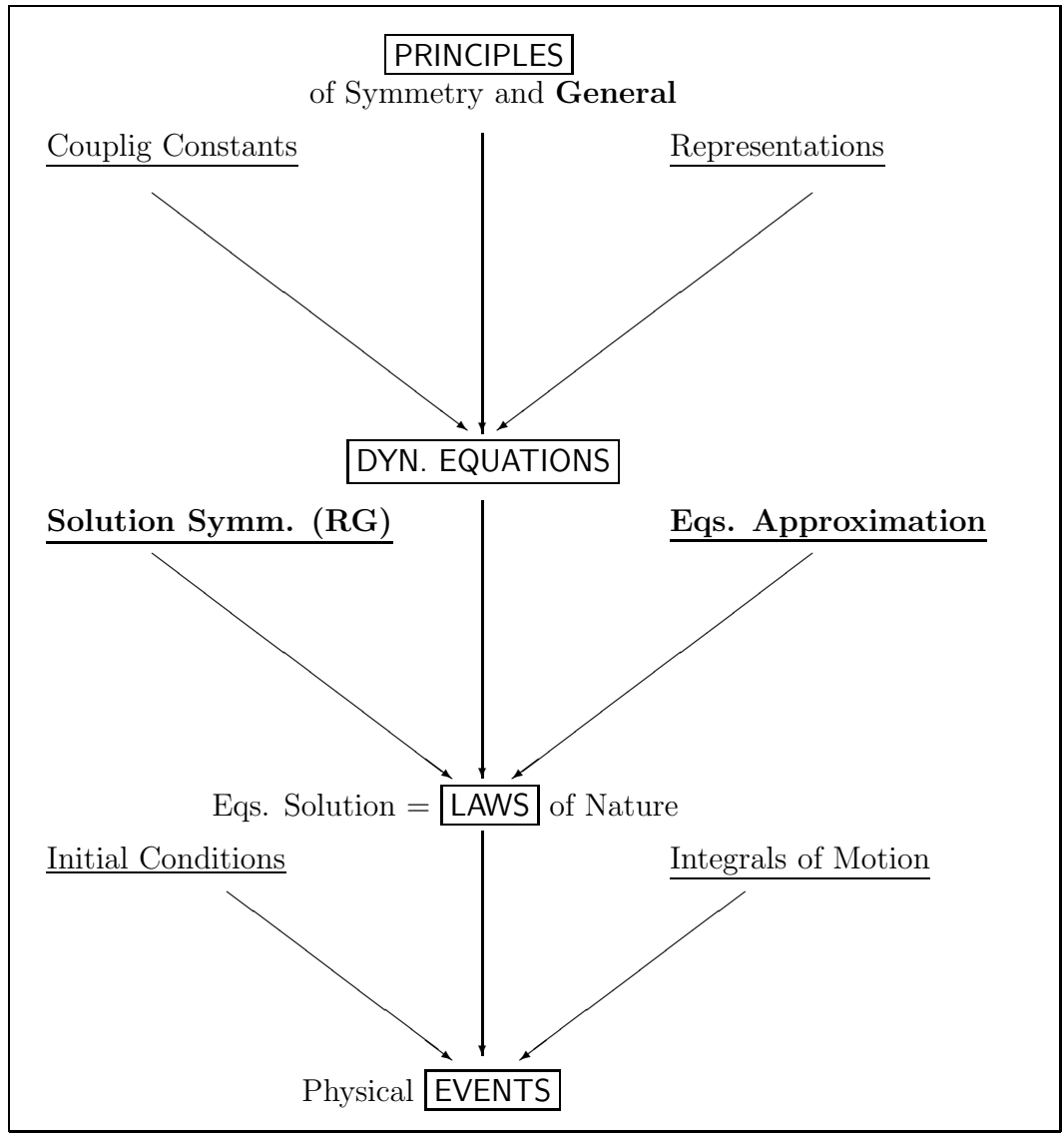
\begin{figure}[h]
\noindent{\large\bf 4. "Logic of Modern Science" Scheme} \\
\vspace{3mm}
        \begin{center}
\begin{picture}(140,150)
\put(0,0){\line(1,0){140}}
\put(0,150){\line(1,0){140}}
\put(0,0){\line(0,1){150}}
\put(140,0){\line(0,1){150}}
\put(50,142){\fbox{\sf PRINCIPLES} }
\put(40,137){of Symmetry and {\bf General} }
\put(5,130){\underline{Couplig Constants}}
\put(88,130){\underline{Representations}}
\put(45,88){\fbox{\sf DYN. EQUATIONS} }
\put(63,133){\vector(0,-1){40}}
\put(19,125){\vector(4,-3){42}}
\put(107,125){\vector(-4,-3){42}}
\put(5,80){\underline{\bf Solution Symm. (RG)}}
\put(88,80){\underline{\bf Eqs. Approximation}}
\put(27,42){Eqs. Solution = \fbox{\sf LAWS} of Nature }
\put(63,84){\vector(0,-1){37}}
\put(19,77){\vector(4,-3){40}}
\put(107,77){\vector(-4,-3){40}}
\put(5,35){\underline{Initial Conditions} }
\put(88,35){\underline{Integrals of Motion}}
\put(39,4){Physical \fbox{\sf EVENTS} }
\put(63,40){\vector(0,-1){30}}
\put(29,32){\vector(4,-3){30}}
\put(98,32){\vector(-4,-3){30}}
\end{picture}
          \end{center}
	 \caption{
	 {\bf The Logic of Modern Science:}
	 {\sl General and Symmetry Principles provide a structure
	 to Dynamic Equations.
	 Solution to Eqs. give the Laws.
			        }     }
	  \label{fig4}
          \end{figure}

\newpage
\noindent{\large\bf 5. Reductionism vs Constuctivism} \\

 We believe that this modification has a direct relation to the
debate between reductionists and constructivists. Two different
credos have been formulated by Einstein (see, e.g. in
Ref.~\cite{einstein})
\bigskip

        \centerline{\underline{ EINSTEIN reductionism}}

\medskip

"The supereme test of the physicist is to arrive at those universal
 elementary laws from which the cosmos can be built up by pure deduction.
\hfill (1918) \par
"...we would like not only to know how nature is organized (and how
 natural phenomena proceed), but possibly to achieve the goal which
 may be considered as utopian and daring -- understand why nature is
 just the way it is". \hfill (1929)  \par
    \bigskip
\noindent and P.~Anderson~\cite{anderson}

        \centerline{\underline{ Modern constructivism}}
\vspace{1mm}
    "The ability to reduce everything to simple fundamental laws does
not imply to start from these laws and reconstruct the Universe"

    "...the more the elementary-particle physicists tell us about
the nature of the fundamental laws, the less relevance they seem
to have to the very real problems of the rest of science..."

\vspace{1mm}

  In our opinion, the real origin of constuctivists' scepticism is just
the gap between "fundamental laws", that is implementation of Principles
in the form of "Equations" (like Newton's, Maxwell's and QCD equations)
and "Laws of Nature" that is solution to the equations (like Kepler's,
Ohm's, Meissner's and Confinement Laws). The RG plays an important
role in this gap filling.\\
\vspace{1mm}

\noindent{\large\bf 6. Renormalization Group -- Solution Symmetry} \\

   Renormalization group first discovered in QFT by Stueckelberg and
Petermann was explicitly formulated by Bogoliubov and the present
author \cite{bsh1} as an exact \cite{gml} group of transformations
related to finite Dyson's transformations. Later on, it has been shown
 \cite{fss} that this exact
group (which we call \cite{umn} the {\sf Bogoliubov Renorm-Group})
is related to the symmetry of a given solution and consists of
specific transformations of a scale and solution parameter(s)
(that could involve, e.g., boundary condition parameters, like
experimentally measured coupling constants); in a particular case
this symmetry can be reduced to power
self-similarity symmetry well known in mathematical physics. \par
    The {\sf Renormalization Group Method}(RGM) devised in
Refs.~\cite{bsh1} (see also English publications~\cite{bsh2}) allows
one to improve an approximate solution
behaviour in the vicinity of a singularity by restoring the correct
structure of this singularity.  \par
    As it is well known, the RGM proved to be an indispensable tool for
ana\-lysing solution property of complicated nonlinear problems in :
 QFT (Ghost problem in QED; asymptotic freedom in QCD; Standard Model
and Grand Unification), critical phenomena and phase transitions,
 percolation, turbulence, polymer theory and many others (including
boundary value problems of mathematical physics~\cite{venia}.  \par
  In this context, we conclude that RG Symmetry being the property of a
 solution forms the basis for "filling a gap between equation and
its explicit solution", solution that is necessary for the 'physical
law' obtaining. \\
 \par
The author would like to thank Prof. A.M.~Baldin for interest in the work
and useful comments. Partial support of RFBR 96-15-96030 grant is gratefully
acknowledged.              \\
\vspace{1mm}   				    


\begin{thebibliography}{50}
\bibitem{bcs} {\small Note here, that the famous BCS model is a
            semi-phenomenological one as it is based upon an effective
            Hamiltonian with direct electron--electron attraction which
            has been introduced 'by hands'. Only the Bogoliubov theory
            of superconductivity devised (independently and practically
            simultaneously) on the basis of Fr\"ohlich mechanism of
            electron--phonon interaction is a true microscopical one.}
\bibitem{gml} {\small As distinct from an approximate group that can be
            connected with short-distant analysis by Gell-Mann and Low or
            approximate semi-group known as "Wilson RG"}.
\end{thebibliography}
\end{document}